\documentclass[global,twocolumn,referee]{svjour}
\usepackage{graphics}
\usepackage{epsfig}
\usepackage[sort&compress,numbers]{natbib}
\journalname{Rheologica Acta}
\begin{document}
\title{Shear localisation with 2D Viscous Froth and its relation to the Continuum Model}
\author{Joseph D. Barry \and Denis Weaire \and Stefan Hutzler% etc
}                     % Do not remove
%
%\offprints{Joseph D. Barry}          % Insert a name or remove this line
\mail{barryjo@tcd.ie}
\institute{School of Physics, Trinity College Dublin, Ireland}
\date{Received: date / Revised version: date}
\maketitle
\begin{abstract}
Simulations of monodisperse and polydisperse ($\mu_2(A)=0.13\pm0.002$) 2D foam samples undergoing simple shear are performed using the 2D Viscous Froth (VF) Model. These simulations clearly demonstrate shear localisation. The dependence of localisation length on the product $\lambda V$ (shearing velocity $V$ times external wall friction coefficient $\lambda$) is examined and is shown to agree qualitatively with other published experimental data. A wide range of localisation lengths is found at low $\lambda V$, an effect which is attributed to the existence of distinct yield and limit stresses. The general Continuum Model is extended to incorporate such an effect and its parameters are subsequently related to those of the VF Model. A Herschel-Bulkley exponent of $a=0.3$ is shown to accurately describe the observed behaviour. The localisation length is found to be independent of $\lambda V$ for monodisperse foam samples. 

Paper presented at 5th Annual European Rheology Conference (AERC), April 15-17, 2009, Cardiff, United Kingdom.
\end{abstract}
%We speculate on the source of localisation in our VF simulations as we approach the quasi-static limit. Finally, we provide a definition of the critical cross-over point into the quasi-static regime
%
\section{Introduction}
\label{intro}

Foam is defined as a two-phase system in which a dispersed phase of gas is enclosed by a continuous phase of liquid \cite{foamsbook}. In aqueous foams, the dispersed phase is typically air, and the liquid phase water with an added surfactant. To simplify the task of understanding the rheology of these systems, it has become popular to concentrate on two-dimensional (2D) foams, which consist of a single planar layer of bubbles. It is the rheology of these systems which is under scrutiny in this paper, using the 2D VF Model. 

We endeavour to understand the mechanisms which cause \emph{localisation of shear} at a moving boundary, as reported in a number of recent experiments (see below). This behaviour is observed in the VF simulations that we will discuss. In the literature, the extent of this shear localisation effect is measured by a localisation length. In this paper, two different definitions of localisation length will be employed.

Simulation results for polydisperse froths subjected to simple shear will be presented, where the localisation length is found to vary with $\lambda V$ (viscous drag times driving velocity; why this is the important parameter to consider is explained in Sec. \ref{sec:details}). For low $\lambda V$, a wide range of localisation lengths is found; see Sec. \ref{sec:details}. This effect is attributed to the existence of distinct yield and limit stresses. We give qualitative evidence of this in Sec. \ref{sec:Continuum}, where we extend the general Continuum Model to incorporate such an effect. In Sec. \ref{sec:multiscale} we proceed to relate parameters of the VF Model to the Continuum Model which is shown to accurately predict the observed behaviour for a Herschel-Bulkley exponent of $a=0.3$. The localisation length is found to be independent of $\lambda V$ in the monodisperse case.

%In particular, we show that for dry, ordered froths subjected to simple shear, the width of the flowing region is independent of $\lambda V$
%A Herschel-Bulkley (HB) exponent of $a=0.32$ is found by relating our simulation data to the Continuum Model theory. We speculate on the source of localisation in our simulations and provide a definition of the critical cross-over point into the quasi-static regime.

In experiments with 2D foams, often a single layer of bubbles is confined between two narrowly spaced glass plates (a Hele-Shaw cell). There are also other types of quasi-2D systems, such as the Bragg raft \cite{Bragg:47}, where a single layer of bubbles floats on a liquid pool, and the confined bubble raft, where a Bragg raft is trapped underneath a glass plate. The most important distinction between these experimental realisations is concerned with the presence of viscous drag. When a foam is in contact with one or two confining plates, there is a drag force associated with any movement of the foam relative to the plate(s). As we shall see, the drag force in the VF Model (see eq. \ref{eq:viscous_froth_original}) plays a role in the localisation of flow in our foam samples.

\begin{figure}[htbp]
\begin{center}
\includegraphics[width=9cm,angle=0]{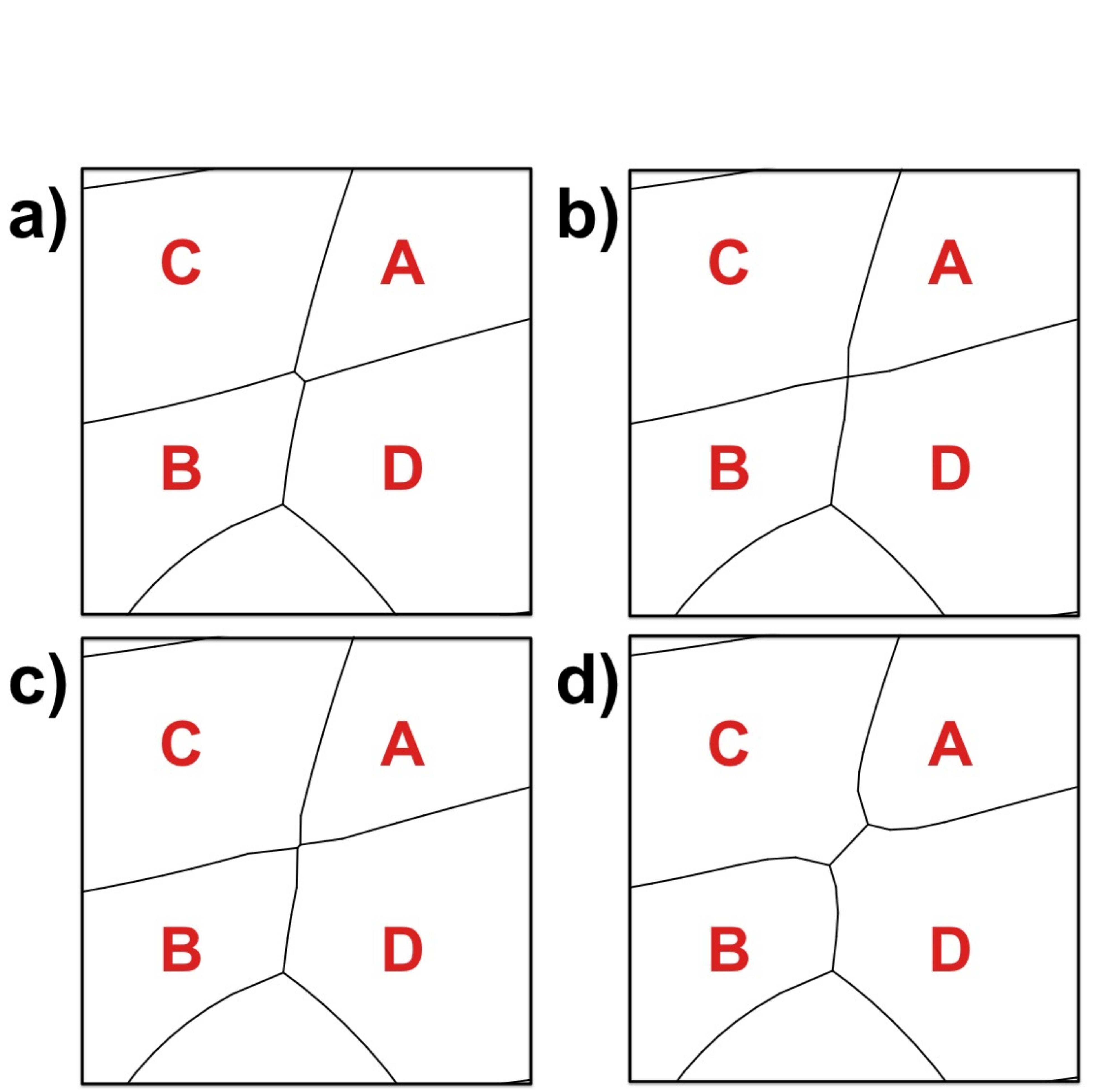}
\end{center}
\caption{A T1 neighbour-swapping event triggered by applying shear. a) Initial configuration, b) A and B lose their common edge, creating an unstable four-fold vertex point, c) a new edge is created between C and D and d) final configuration. Data taken from a VF simulation.}
\label{f:T1}
\end{figure}

When a 2D foam is subjected to an applied shear stress, after an initial transient, it yields and begins to flow. The foam yields locally when the \emph{yield stress} is reached. A more detailed description of the stress-strain relation will be required when interpreting the simulation results presented in this paper; see Sec. \ref{sec:Continuum}. At the local level, yielding is due to plastic events, i.e. T1 topological changes of the foam structure (see Fig. \ref{f:T1}). Two neighbouring bubbles (A and B) lose a common edge which is subsequently gained by two proximate bubbles (C and D), which become neighbours. We describe the incorporation of these topological changes into the VF Model in Sec. \ref{sec:vf}.

When flow is concentrated in one region and not in another, the flow is said to have localised. Debreg\'{e}as \emph{et al.} \cite{Debregeas:01} were the first to report definitive evidence of shear localisation in 2D aqueous foams. Their experiments exhibit shear localisation next to a moving boundary in a Couette geometry, with an exponential decay in the measured foam velocity profiles. Similar results have been reported by Wang \emph{et al.} \cite{Wang:06} and Krishan and Dennin \cite{Krishan:08} for straight and circular geometries, respectively. These results have been interpreted within the framework of the Continuum Model \cite{Janiaud:06b,Janiaud:07,Clancy:06} where shear localisation is attributed to the presence of a drag force. This notion is further supported by the work of \cite{Green:09} which studies the effects of drag forces at high shear rates using the VF Model. However there are also quasi-static simulations showing localisation (discussed below) in which there is no such wall drag. This suggests that there is more than one mechanism that may lead to shear localisation. We will return to this point in Sec. \ref{sec:Continuum}.

% in a hexagonal honeycomb staircase structure undergoing fast shear

Experimental work by Katgert \emph{et al.} \cite{Katgert:08,Katgert:09} on the shearing of bidisperse foams in a Hele-Shaw cell (straight geometry, that is, simple shear) shows Herschel-Bulkley behaviour (discussed below) and supplies further evidence of shear localisation in 2D foams. In this case, however, the velocity profiles are not exponential. Such non-exponential velocity profiles, together with their velocity dependence can be obtained from an extension or generalisation of the original Continuum Model \cite{Weaire:08a,Weaire:09}. Furthermore, these experiments show that the localisation length decreases as the velocity of the moving boundary increases. In this paper, we show velocity profiles from our VF simulations which exhibit qualitatively similar behaviour.

Of particular current interest in these types of experiments is the dependence of (local) shear stress on shear rate. This effect is captured by the Herschel-Bulkley constitutive relation,
\begin{equation}
\label{eq:hb-relation}
\sigma=\sigma_{y}+c_{v}\dot{\epsilon}^a
\end{equation}
\noindent
where $\sigma$ is stress, $\sigma_{y}$ is the yield stress, the coefficient $c_{v}$ is the so-called consistency, $\dot{\epsilon}$ is strain rate and $a$ is the HB exponent. Katgert and co-workers report $a=0.36$. They also note that in the monodisperse case (i.e. bubbles of equal size), the localisation length is found to be independent of shear rate. We too find this to be the case in our simulations; see Sec. \ref{sec:details}.

Shear localisation has been studied computationally using other microscopic (bubble scale) models. Quasi-static models, as explored by \cite{Weaire:83,Weaire:84,Bolton:90,Hutzler:95} might shed light on behaviour at very low strain rates. Results reported by Kabla \cite{Kabla:07a,Kabla:07b} show localisation next to the boundaries in quasi-static shearing simulations ($\mu_2(A)\approx0.06$ using the definition eq. \ref{eq:disorder}), consistent with experimental observation. In these simulations of simple shear, where there is no wall drag, either wall may be regarded as the one that moves. Recent results by Wyn \cite{Wyn:08} suggest that as the second moment of the bubble area distribution $\mu_{2}(A)$ is increased in such simulations to values approaching 0.2 or higher, shear-banding can occur in regions away from the moving boundary. The width of these shear bands has a square-root dependence on $\mu_{2}(A)$. This type of behaviour has yet to be observed in experiments. In this paper, we only examine the rheology of foam samples of disorder $\mu_2(A)=0.13$ but will probe higher values of disorder in future work in order to search for similar effects.

With the aid of the Surface Evolver software \cite{Brakke:96}, quasi-static simulations are increasingly easy to implement and, with modern computing, are certainly fast. But are they suitable for rheology? In quasi-statics, the foam is relaxed to equilibrium at each step. There is therefore no relevant time scale present and so no concept of shear-rate. It makes no sense to consider Herschel-Bulkley type relations or to discuss the dependence of localisation on boundary velocity.

What then, are the alternatives to quasi-static simulations? Bubble models \cite{Durian:95}, where a foam is modeled as a collection of interacting disks appear to represent at some level the dynamics of 2D foams. Langlois \emph{et al.} \cite{Langlois:08} report a Herschel-Bulkley exponent of $a=0.54$ ($\mu_2(A)\approx0.03$). In addition, shear localisation is observed when wall drag is present. For dry foams though, where low liquid fraction causes bubbles to become more polygonal in shape, this approach is no longer accurate \cite{Green:09}. 

In this paper, we adopt the 2D Viscous Froth (VF) Model \cite{Kern:04,Green:06} as a more realistic model for dry 2D foam dynamics. We have performed an extensive study of shear localisation with the VF Model in a straight geometry which shows realistic dynamics and a rich variety of behaviour, particularly at low $\lambda V$.

For a summary of the experimental and theoretical work presented in this section, see \cite{WeaireBarryHutzler:09}.

\section{The 2D Viscous Froth Model and its implementation}
\label{sec:vf}

\begin{figure}[htbp]
\begin{center}
\includegraphics[width=5cm,angle=0]{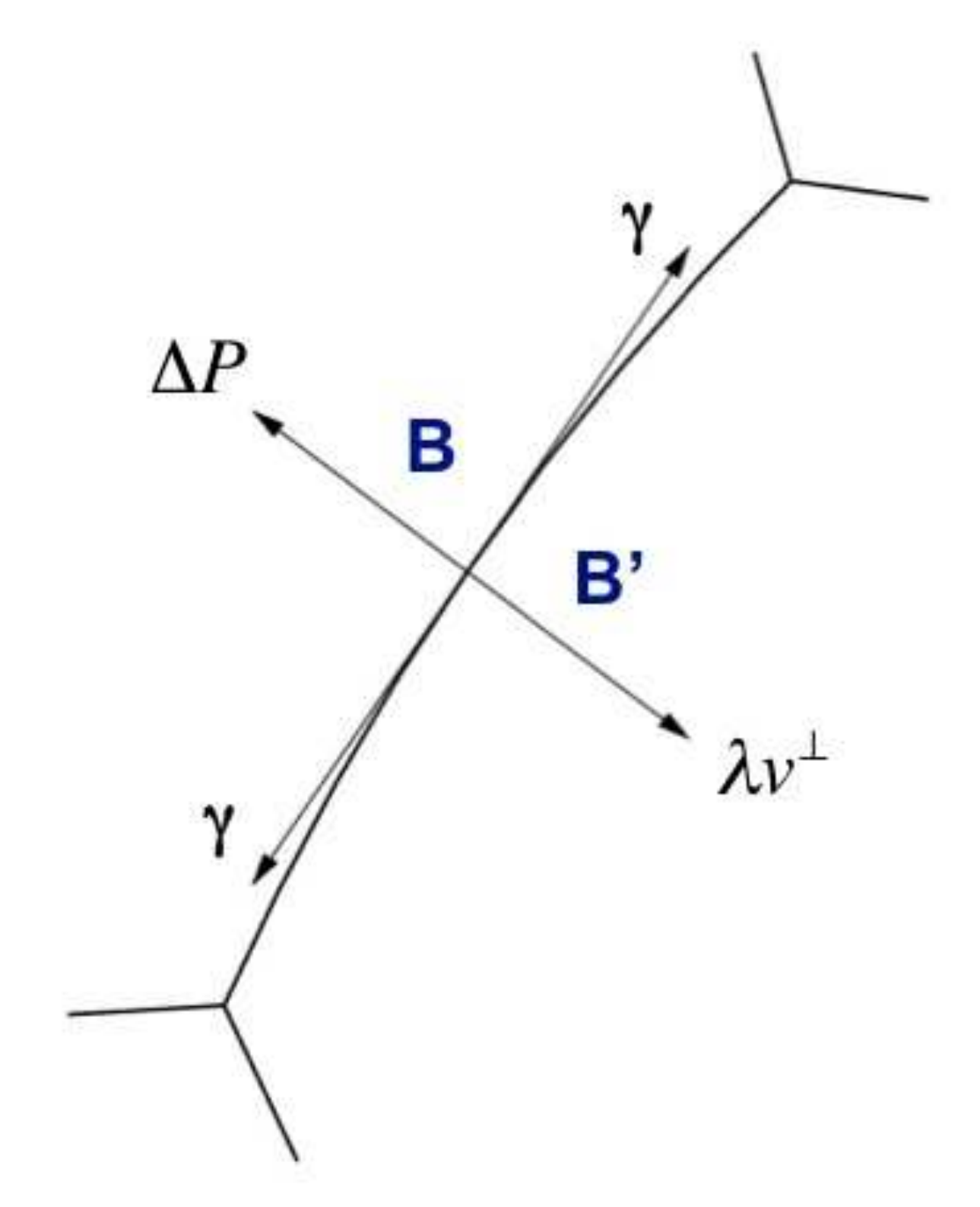}
\end{center}
\caption{A diagram illustrating the various forces involved in film motion with the two-dimensional VF Model. $B$ and $B^\prime$ indicate the two bubbles the central soap film is separating. Note that this film is in contact with a surface (in the plane of the page) which results in a drag force when the film is moving.}
\label{f:VF_forces}
\end{figure}

The model describes the motion of a soap film in the 2D systems described above, with wall drag \cite{Kern:04}. In the present case, bubble areas are kept constant. The foam is considered to be sufficiently dry (liquid fraction less than 0.01) so that a soap film may be accurately described by a curved line and the junctions are represented by points. In the present simulations, a soap film is approximated as a system of connected straight line segments. The motion of a point $s$ joining these segments is given via the equation

\begin{equation}
\lambda v^{\perp}(s)=\Delta P- \gamma K(s)
\label{eq:viscous_froth_original}
\end{equation}
\noindent
where $\lambda$ is the wall drag coefficient, $v^{\perp}(s)$ is the velocity of a point $s$ in the direction of the normal vector $\vec{N}(s)$ to the soap film, $\Delta P$ represents relative pressure differences between neighbouring cells, $\gamma$ is a constant surface tension force (in 2D), and $K(s)$ is local film curvature calculated from the  relative positions of adjacent (discrete) film segment points. See Fig. \ref{f:VF_forces} for an illustration of the forces involved. Setting $\lambda=0$ in eq. \ref{eq:viscous_froth_original} recovers the Young-Laplace law, corresponding to soap films that are arcs of circles.

Throughout the implementation of the model, film segments adjacent to the three-fold vertex points are held an angle of $\frac{2\pi}{3}$ radians relative to each other, in accordance with Plateau's rules for a soap froth. Details on the numerics of this calculation are best found in \cite{Green:06}. It should be noted that for high rates of strain, one would expect surface tensions in the soap films to vary to the point that this equilibrium condition would no longer apply (for example, because of the Marangoni effect). At least for lower rates of strain, the $\frac{2\pi}{3}$ rule is reasonable. 
%We verify that we are in a regime of low strain rate in Sec. \ref{sec:details} by measuring the range of Deborah number (see eq. \ref{eq:deborah}) associated with our VF simulations.

The VF model may be conveniently incorporated into a Surface Evolver \cite{Brakke:96} script (as pioneered by Cox \cite{Cox:05}), thus allowing for the use of various SE features. The procedure for performing (T1) topological changes in the Surface Evolver is as follows, and is illustrated in Fig. \ref{f:T1}. The distance along the film between neighbouring three-fold vertex points is calculated at each timestep.  When this film length becomes smaller than a predefined critical cut-off length, $l_{c}$ then it is deleted using the Evolver's `edgeweed' command. A four-fold vertex is temporarily formed to maintain the topology of neighbouring cells (see Fig. \ref{f:T1}(b)). The Evolver's `pop' command is then employed, which scans the foam for vertices which do not have a legal topology and replaces the four-fold vertex with a proper local topology. This results in a new film of effectively negligible length oriented in the perpendicular direction to the old film (see Fig. \ref{f:T1}(c)). 

Further details on the implementation of the VF Model can be found in the papers by Kern \emph{et al.} \cite{Kern:04} and Green \emph{et al.} \cite{Green:06}.

\section{Sample Creation}
\label{sec:creation}

\begin{figure*}[htbp]
\begin{center}
\includegraphics[width=14cm]{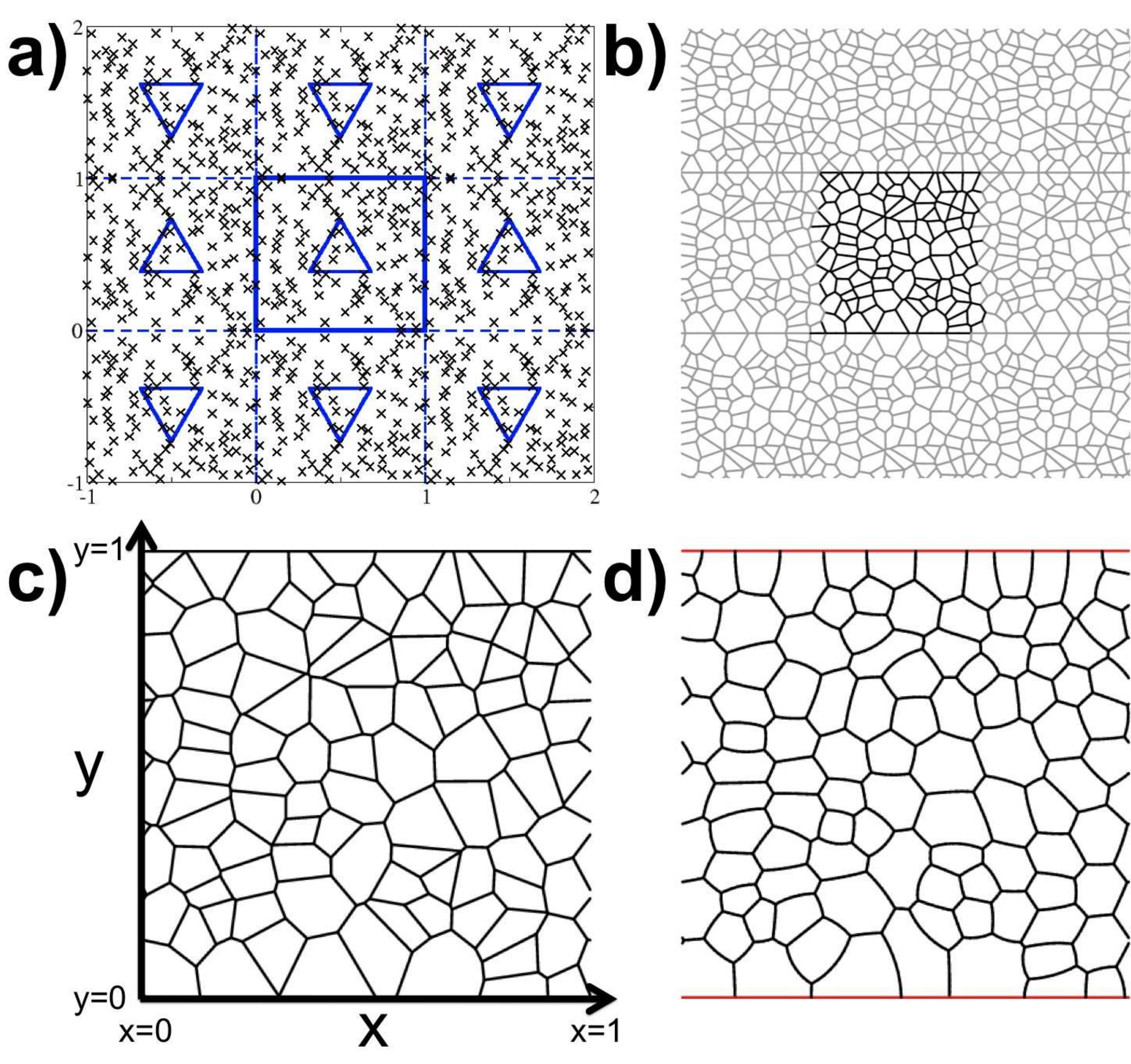} \hfil
\end{center}
\caption{The creation of a semi-periodic polydisperse two-dimensional foam sample as required for our simple shear simulations. a) Points are placed in the central unit cell and translated/reflected into adjacent boxes (as indicated by the background triangles). b) The Voronoi Diagram of these points is calculated. c) The central area is isolated and made into a half-periodic (in x-direction) data-file. d) Keeping cell areas constant, the Surface Evolver performs line minimisation on the structure. We refer to this shown structure as Sample 1 later in the text.}
\label{f:sample_creation}
\end{figure*}

A semi-periodic monodisperse sample is created using the standard method outlined in the Surface Evolver documentation (which is supplied with the software package). Disordered semi-periodic samples are created by the following process (illustrated in Fig. \ref{f:sample_creation}).

%Points are placed in the unit cell according to an algorithm which allows us to control the minimum distance between them, see the central box in Fig. \ref{f:sample_creation}(a). 

Points are placed at random in the unit cell using a uniform distribution to determine the x and y positions; see Fig. \ref{f:sample_creation}(a). New points are added to the box if they are more than a predefined minimum distance $r_{min}$ from any other point. The process is continued until the desired number of points have been successfully placed. A lower value of $r_{min}$ results in more polydisperse samples. These points are translated to boxes to the left and right, and reflected (see Fig. \ref{f:sample_creation}(a)) through the lines $y=0$ and $y=1$ to boxes above and below, based on the method of De Fabritiis and Coveney \cite{Fabritiis:03} (as indicated by the background triangles in Fig. \ref{f:sample_creation}(a)). With the software package Qhull \cite{qhull:96}, the Voronoi Diagram  (a particular way of tessellating the plane into regions of convex polygons) of these points is calculated; see Fig. \ref{f:sample_creation}(b). This is then passed into the Surface Evolver. The box in the centre is isolated (see Fig. \ref{f:sample_creation}(c)) and keeping the areas of each of the cells fixed, the Surface Evolver performs line minimization on the structure. The resulting structure is our final two-dimensional half-periodic (i.e. periodic in the x-direction only) foam data file; see Fig. \ref{f:sample_creation}(d). Of interest here (as a result of the reflection) is that the straight line boundaries at $y=0$ and $y=1$ naturally occur as a result of this process.

\section{Simulation Details and Results}
\label{sec:details}

Using the above methods for foam sample creation, we create one monodisperse foam sample and five foam samples of polydispersity $\mu_{2}(A)=0.13\pm0.002$, where the measure of polydispersity is defined as the second moment of the area distribution,

\begin{equation}
\label{eq:disorder}
\mu_2(A)=\overline{\left(1-\frac{A}{\bar{A}}\right)^2}\ .
\end{equation}
\noindent
Here $A$ denotes the area of a bubble, and $\bar{A}$ the mean bubble area.

The foam samples consist of $N_{b}=100$ bubbles in a square unit cell of area 1; it is too computationally expensive to run larger samples in a VF simulation. In our dimensionless simulation units, our system size $L=1$ and mean bubble area $\bar{A}=0.01$. We define a new length scale ${\bar{A}}^{1/2}$, the square root of the mean bubble area. In these new units, $L=10\ {\bar{A}}^{1/2}$. The width $W_{l}$ of one layer of bubbles in our square sample is given by
\begin{equation}
\label{eq:layer-width}
W_{l}=L / \sqrt{N_{b}}={\bar{A}}^{1/2}
\end{equation}

We proceed to move the top boundary in the positive x-direction with velocity $V$ by incrementally moving vertices at $y=1$ a distance $Vdt$ per timestep $dt$. The VF algorithm, as outlined in Sec. \ref{sec:vf} is used to determine the dynamics of the foam during each timestep. Typical values for the displacement of the shearing boundary per timestep are in the range $10^{-6}{\bar{A}}^{1/2} \le Vdt \le 10^{-3}{\bar{A}}^{1/2}$ (depending on what values of $V$ and $\lambda$ are used). No-slip boundary conditions are maintained by fixing vertices lying on the boundaries while the VF algorithm is being implemented. Our boundary conditions are thus

\begin{equation}
\left\{
\begin{array}{c}
v(L)=V \\
v(0)=0
\end{array}
\right.
\label{eq:bc}
\end{equation}

%The value of surface tension $\gamma$ is set to 1, and
Multiple simulations are run for different values of  $\lambda V$ (wall drag coefficient times boundary velocity) with a fixed value of surface tension $\gamma$. To see why this is the appropriate parameter to look at, consider again the equation of motion for the VF Model, as given by eq. \ref{eq:viscous_froth_original}. By setting $v^\perp=V\hat{v}^\perp$, where $V$ is the boundary velocity and $\hat{v}^\perp$ is our rescaled dimensionless velocity, we can rewrite our equation of motion as

\begin{equation}
\label{eq:scaling}
(\lambda V)\hat{v}^\perp(s)=\Delta P- \gamma K(s)
\end{equation}
\noindent
It is clear that, given any initial state configuration, its development in time is determined by $\lambda V$. Furthermore, we see evidence of this $\lambda V$ dependence if we rewrite the Herschel-Bulkley relation (see eq. \ref{eq:hb-relation}) in terms of our VF parameters. As stress in 2D has dimensions of force per length, on dimensional grounds, we see that

\begin{equation}
\sigma=\sigma_y+\hat{c_v}\gamma^{1-a}{\bar{A}}^{a-1/2}L^{-a}(\lambda V)^a
\label{eq:viscous-stress}
\end{equation}
\noindent
where the 2D surface tension $\gamma$ has dimensions of force, $\lambda V$ has dimensions of force per length and $\hat{c_v}$ is a dimensionless parameter of order unity which may be related to $\mu_2(A)$. In this derivation, we define the strain rate term of eq. \ref{eq:hb-relation} as the nominal shear rate of the system, $\dot{\epsilon}=V/L$.

%We will also encounter this product of $\lambda V$ in Sec. [XXX] when re-writing the Herschel-Bulkley relation in terms of the parameters of the Viscous Froth Model.

%We shall see later that the calculated localisation lengths of our Viscous Froth simulations are a function of the product $\lambda V$. We arrive at this conclusion using dimensional arguments in Sec. [X].

%The data of Fig. \ref{f:scaling} was used to check this conclusion.

\begin{figure}[htbp]
\begin{center}
\includegraphics[width=9cm]{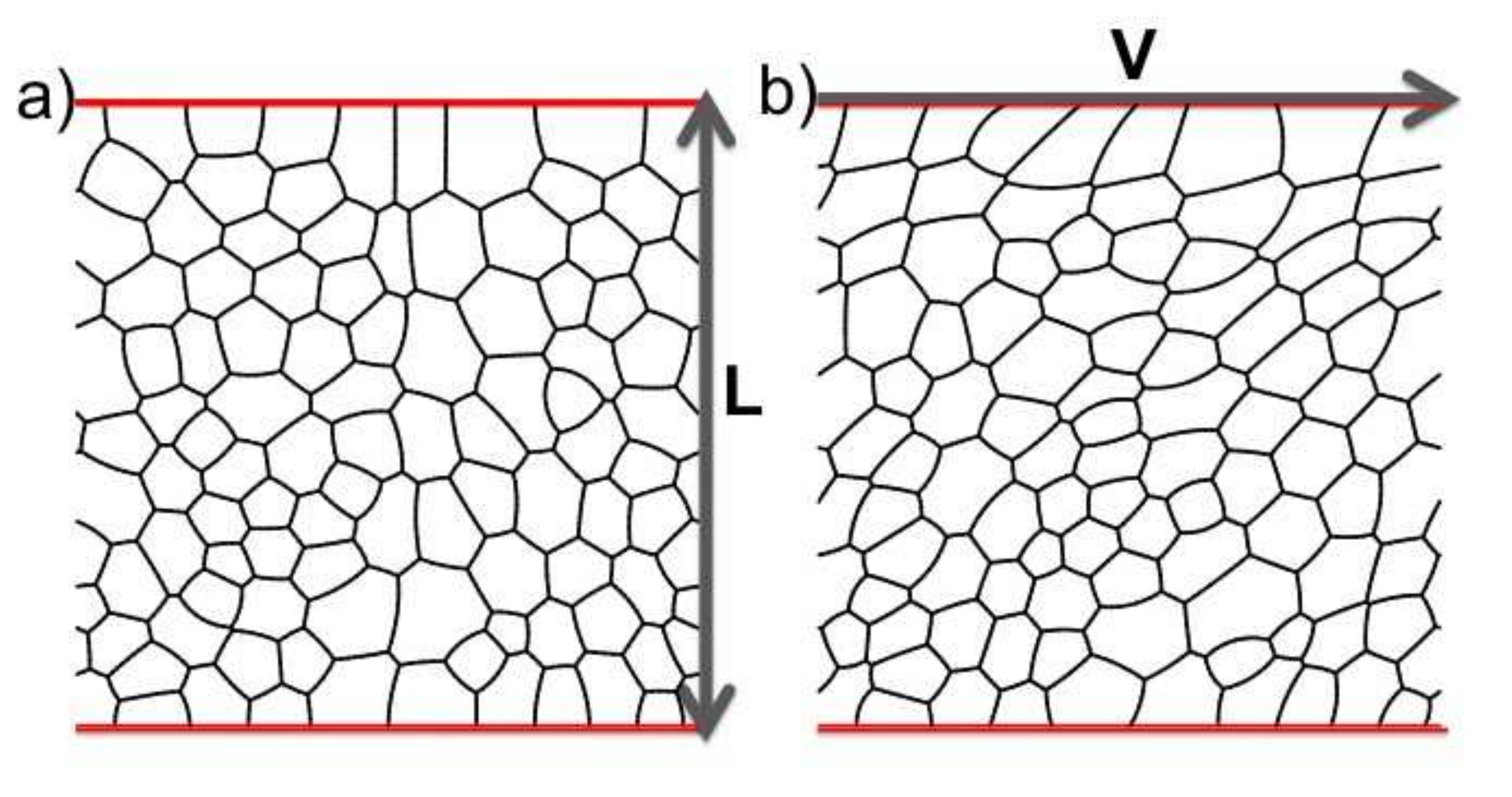} \hfil
\end{center}
\caption{(a) A polydisperse foam consisting of 100 bubbles (Sample 5) in equilibrium. $L$ denotes our system size. (b) The same foam being sheared at a velocity $V$.}
\label{f:poly-setup}
\end{figure}

To calculate flow profiles, bubble centre positions are determined. We subsequently divide our foam into bins of width $W_{l}$ and calculate the average velocity of bubbles centres in each bin over time. A sketch of our simulation setup is illustrated in Fig. \ref{f:poly-setup} (polydisperse sample).

\begin{figure}[htbp]
\begin{center}
\includegraphics[width=9cm]{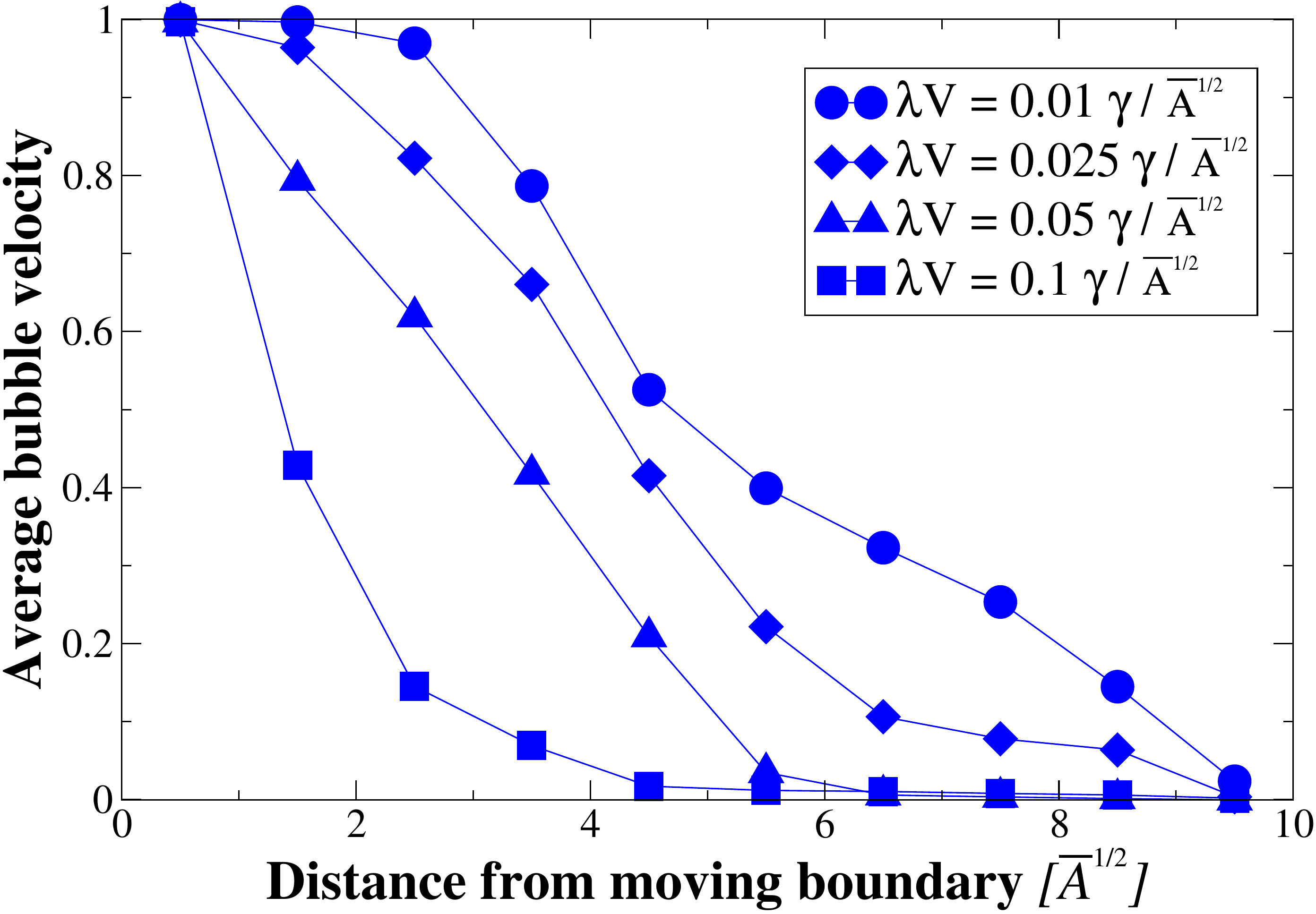} \hfil
\end{center}
\caption{Examples of velocity profiles for different values of $\lambda V$. Profiles shown are for Sample 1. The corresponding localisation lengths for these profiles are denoted by filled circles in Fig. \ref{f:poly-localisation-velocity}.}
\label{f:poly-profile-drag}
\end{figure}

\begin{figure}[htbp]
\begin{center}
\includegraphics[width=9cm]{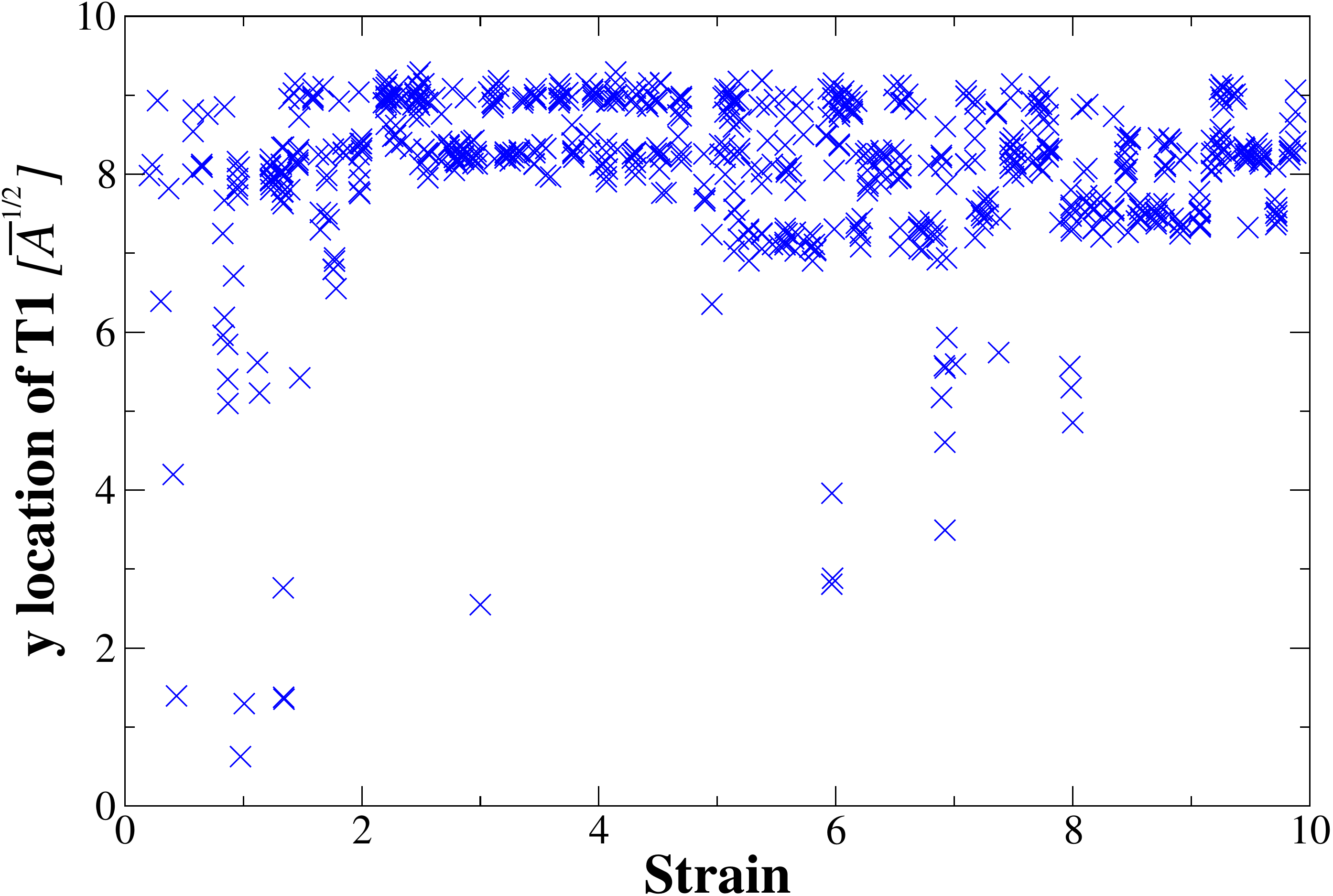} \hfil
\end{center}
\caption{The location of T1 topological changes as a function of strain for a polydisperse sample. After an initial transient, where T1s happen everywhere in the foam, the flow localises and T1s are found to occur mostly next to the moving boundary (at $y=10\ \bar{A}^{1/2}$). Shown data is for Sample 1 where $\lambda V=0.05\ \gamma / \bar{A}^{1/2}$}.
\label{f:poly-t1}
\end{figure}

Fig. \ref{f:poly-profile-drag} shows examples of averaged steady state velocity profiles. We say that a simulation has reached a steady state once there is no longer any appreciable change in our velocity profile in time. Typically, we average our steady state velocity profiles over the range $1\le \epsilon \le10$, where the imposed strain $\epsilon$ is defined as $\epsilon=\Delta x / L$ and $\Delta x$ is equal to the total displacement of the moving boundary. Note that there is a clear change in the flow profiles as we vary $\lambda V$. We find that localisation occurs close to the moving boundary in all but two of our simulations. (In one of these cases, for $\lambda V=0.01\ \gamma\bar{A}^{1/2}$, localisation switches to the stationary boundary, while in the second case, for $\lambda V=0.005\ \gamma\bar{A}^{1/2}$, a shear band occurs in the centre of the sample away from either boundary (data not shown).)  Localisation of flow can also be made visible by plotting the positions at which T1 topological changes occur in our samples, as done by \cite{Wyn:08}. An example is shown in Fig. \ref{f:poly-t1}.

%\begin{figure}[htbp]
%\begin{center}
%\includegraphics[width=9cm]{Profiles_with_fits_velocity2.eps} \hfil
%\end{center}
%\caption{Examples of velocity profiles for different values of boundary velocity $V$ and $\lambda$ set to one. Profiles shown are for Sample 2. The corresponding localisation lengths for these profiles are denoted by open triangles in Fig. \ref{f:poly-localisation-velocity}.}
%\label{f:poly-profile-velocity}
%\end{figure}

At this stage, it is unclear what the form of the velocity profiles is. We have attempted to use exponential fits and fits from the general Continuum Model \cite{Weaire:09} in the data fitting process but this approach does not yield consistently good fits to our velocity profiles which are clearly quite noisy, presumably due to the small system size. To obtain a measure of the width of the flowing region from these noisy profiles, we use the following definition of localisation length, denoted by $l_{int}$ \cite{WeaireBarryHutzler:09}

\begin{equation}
l_{int} = \frac{1}{V}\int_0^L v(y) dy\ .
\label{eq:cont-localisation}
\end{equation}

\noindent

\begin{figure}[htbp]
\begin{center}
\includegraphics[width=9cm]{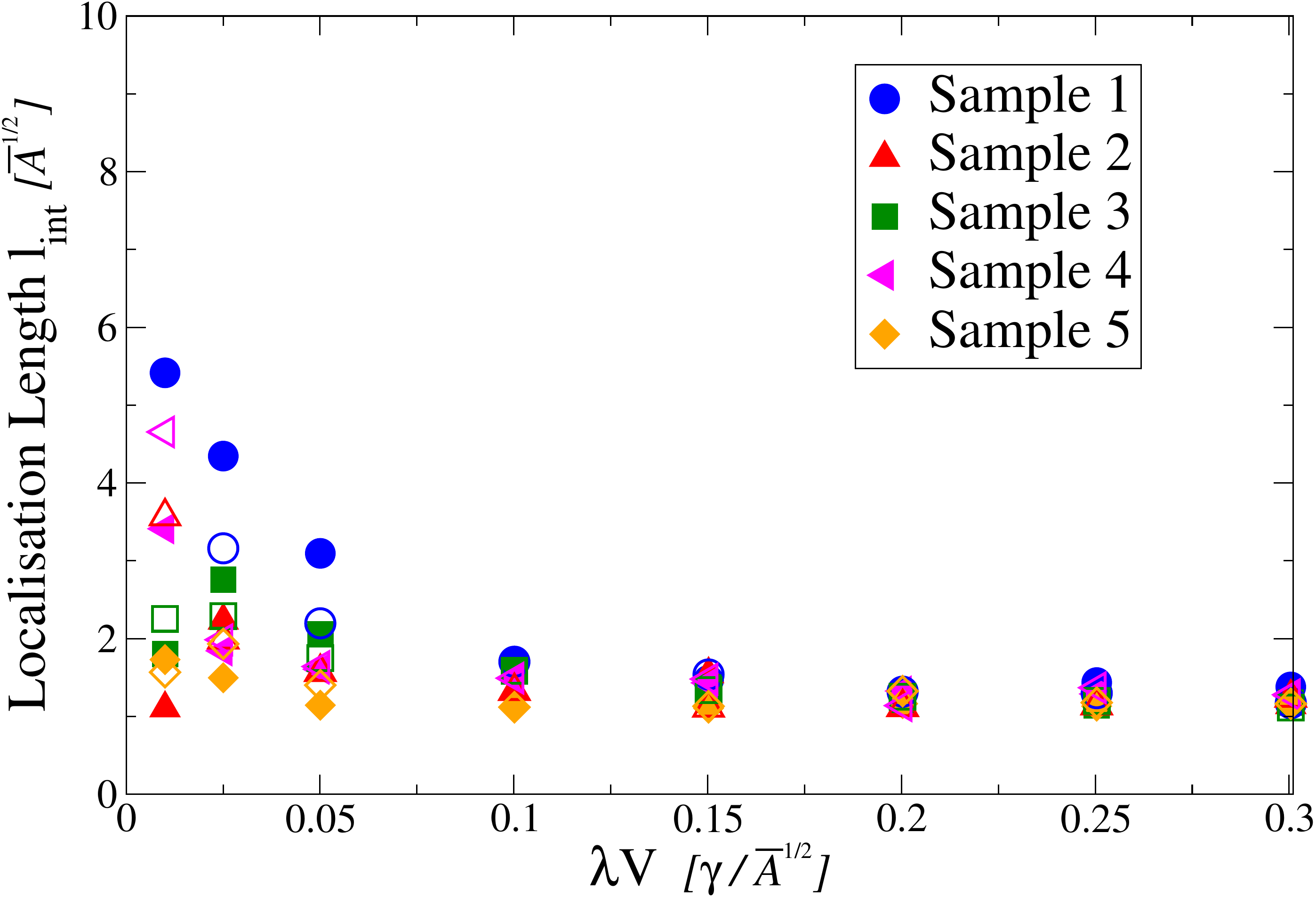} \hfil
\end{center}
\caption{Localisation lengths for a range of $\lambda V$. Each symbol represents one simulation run. For low $\lambda V$, there is a wide range of lengths, while for high $\lambda V$, only the first layer of bubbles flows. Filled symbols indicate simulations where $V$ is fixed and $\lambda$ is varied. Open symbols indicate simulations where $\lambda$ is fixed and $V$ is varied.}
\label{f:poly-localisation-velocity}
\end{figure}

%\begin{figure}[htbp]
%\begin{center}
%\includegraphics[width=9cm]{Localisation_Comparison_Between_Samples_Drag2.eps} \hfil
%\end{center}
%\caption{Localisation length for a range of wall drag $\lambda$ and $V$ set to one. For low drag, there is a wide range of lengths, while for high drag, all curves converge to $l=1.33D\pm0.44D$.}
%\label{f:poly-localisation-drag}
%\end{figure}

This integral, which has the required dimensions of length, is calculated numerically for each of our velocity profiles using the Trapezoidal Rule. Fig. \ref{f:poly-localisation-velocity} shows a variation of localisation length with  $\lambda V$. Note that for low $\lambda V$ we find large scatter in the localisation lengths, however, this scatter decreases as $\lambda V$ is increased. For high $\lambda V$, the length converges towards the minimum localisation length, $l_{min}={\bar{A}}^{1/2}$, the width of one bubble layer  (see eq. \ref{eq:layer-width}). This is because the first layer of bubbles always flows.

The ratio of the intrinsic timescale of the VF Model to the external timescale (as imposed by the nominal shear rate $\dot{\epsilon}=V/L$), otherwise known as the Deborah number $D_e$, is given by

\begin{equation}
D_e=\frac{(\lambda V)\bar{A}}{\gamma L}\ ,
\label{eq:deborah}
\end{equation}
\noindent
as defined in  \cite{Kern:04}. A small Deborah number ($D_e\ll1$) indicates that the foam has enough time available to re-equilibrate, even as the applied shear attempts to bring the foam out of equilibrium (and vice versa for large $D_e$). In our simulations, $0.001\le D_e \le 0.03$. We therefore conclude that we are close to the quasi-static regime in all of the discussed simulations.
%where the foams behave `like a fluid'.

While not shown here, similar simulation runs have been performed for a monodisperse foam ($\mu_{2}(A)=0$). The localisation length is found to be independent of $\lambda V$ and is determined to be $l={\bar{A}}^{1/2}$ (the same as $l_{min}$ in our polydisperse simulations).

Our simulation results are broadly consistent with the findings of Katgert \emph{et al.} \cite{Katgert:08,Katgert:09}, where the localisation length is found to decrease with increasing $V$, and rate independence of localisation length is found in the monodisperse case. 

%However, to understand the source of the observed range of localisation lengths at low $\lambda V$ and to obtain a measure of the important Herschel-Bulkley exponent from these simulation results, we must now turn to the Continuum Model.

We wish to gain an understanding of these VF simulation results by attempting to capture the observed behaviour by a continuum model. Such a model must include a constitutive relation which relates the local (wall) drag force of the VF model to an averaged drag force in the continuum description. However, this would not be enough to explain the observed simulation results, as (according to the general Continuum Model \cite{Weaire:09}) it would result in a zero localisation length in all cases. Therefore, results suggest that internal dissipation (represented by the shear rate term in the HB relation; see eq. \ref{eq:hb-relation}) should also be included, although it is not clear how this dissipation arises in the VF simulations. We will also appeal to the idea of the existence of a stress overshoot in order to explain the variation of localisation length at low $\lambda V$.

%We will return to this point in Sec. \ref{sec:Continuum}.

%We now proceed to discuss an important scaling relation, and then, with the aid of the Continuum model, to explain the observed flow behaviour, and to extract the important Herschel-Bulkley exponent from these results.

%\section{Scaling relation for the Viscous Froth Model}
%\label{sec:scaling}

%The equation of motion for the Viscous Froth Model is given by eq. \ref{eq:viscous_froth_original}. By setting $v=V\hat{v}$, where $V$ is the boundary velocity and $\hat{v}$ is our rescaled dimensionless velocity, we can rewrite our equation of motion as

%\begin{equation}
%\label{eq:scaling}
%(\lambda V)\hat{v}=\Delta P- \gamma K
%\end{equation}
%\noindent
%It is clear that, given any initial state configuration, its development in time is determined by $\lambda V$, and in particular that the localisation length is a function of the product $\lambda V$. The data of Fig. \ref{f:scaling} was used to check this conclusion.

%\begin{figure}[htbp]
%\begin{center}
%\includegraphics[width=9cm]{Avg_DV_separate.eps} \hfil
%\end{center}
%\caption{An average to the localisation lengths shown in Fig. \ref{f:poly-localisation-velocity} ($V$ varying, $\lambda=1$) and Fig. \ref{f:poly-localisation-drag} ($\lambda$ varying, $V=1$), plotted against our control parameter $\lambda V$. The behaviour of both curves is similar, as one would expect from eq. \ref{eq:scaling}.}
%\label{f:scaling}
%\end{figure}

\section{The Continuum Model}
\label{sec:Continuum}

Up until now, we have discussed microscopic (bubble scale) models of 2D foam rheology. An alternative way of describing a foam is to treat it as a continuum. The generalised Continuum Model \cite{Weaire:08a,Weaire:09} (for steady shear) combines the Herschel-Bulkley constitutive relation (see eq. \ref{eq:hb-relation}) with the following expression for the variation of (wall) drag force $F_d$ per unit area as a function of local velocity $v$

\begin{equation}
F_d=-c_{d}v^b\ ,
\label{eq:drag-force}
\end{equation}

\noindent
where $c_{d}$ is the drag coefficient and $b$ is the drag exponent (the Bretherton law gives $b=\frac{2}{3}$ \cite{Bretherton:61}). These two expressions can be related by a force balance, which leads to the following differential equation \cite{Janiaud:06b,Weaire:08a}

\begin{equation}
\frac{d}{dy}\left| \frac{dv(y)}{dy} \right|^a = -\frac{c_d}{c_v}v(y)^b\ ,
\label{eq:diff-equation}
\end{equation}
\noindent
which can be solved using the boundary conditions $v(0)=V$ and $v(L)=0$. These are equivalent to the boundary conditions imposed in our VF simulations; see eq. \ref{eq:bc} (albeit that here the distance $y$ is measured \emph{downward} from the shearing boundary). 

Upon inspection, it is clear that that a velocity profile of the following form satisfies eq. \ref{eq:diff-equation}, and exhibits flow localisation:

\begin{equation}
 v(y)=V (1-y/y_0)^n
 \label{eq:solution}
\end{equation}
\noindent
where $y_0$ and $n$ may be obtained by inserting eq. \ref{eq:solution} into eq. \ref{eq:diff-equation} and equating prefactors and exponents \cite{WeaireBarryHutzler:09}. This gives

\begin{equation}
y_0 = \frac{1+a}{a-b}\left(\frac{a(1+b)c_vV^{a-b}}{(1+a)c_d}\right)^{\frac{1}{1+a}}
\label{e:value_for_y_0}
\end{equation} 
and
\begin{equation}
n=     \frac{1+a}{a-b}\ .
\label{e:value_for_n}
\end{equation}

Eq. \ref{eq:solution} clearly satisfies the first boundary condition, $v(0)=V$ of eq. \ref{eq:diff-equation}. The second boundary condition is satisfied if we take our sample size $L\to\infty$ \cite{WeaireBarryHutzler:09}. This approach is valid so long as the size of the sample is much greater than the localisation length ($L\gg l$) which may be defined as

\begin{equation}
 l=\left| \frac{V}{\frac{dv(0)}{dy}} \right| \ ,
 \label{eq:loc-length}
\end{equation}
\noindent
an alternative definition to that of the previous section; see eq. \ref{eq:cont-localisation}. Inserting eq. \ref{eq:solution} into eq. \ref{eq:loc-length} leads to the following expression for localisation length as a function of the shearing velocity $V$,
%In the limit of the sample size being much greater than the localisation length ($L>>l$), the following analytical expression for variation of $l$ with $V$ was derived by \cite{Weaire:09}

\begin{equation}
l=\left(\frac{a(1+b)c_{v}}{(1+a)c_{d}}\right)^{\frac{1}{1+a}}V^\frac{a-b}{1+a}\ .
\label{eq:continuum-localisation}
\end{equation}
\noindent
Its relation to the definition of localisation length $l_{int}$ (see eq. \ref{eq:cont-localisation}) is
\begin{equation}
l_{int}/l=(1+a)/(1+2a-b)\ ,
\end{equation}
\noindent
as given by \cite{WeaireBarryHutzler:09}. For $a<b$ (as is the case for our VF simulations; see Sec. \ref{sec:multiscale}), localisation length therefore \emph{decreases} as the shearing velocity $V$ is \emph{increased}. 

The possibility of having a range of localisation lengths at low $V$ (as in Fig. \ref{f:poly-localisation-velocity}) can be accounted for by extending the Continuum Model to incorporate what we will refer to as a stress overshoot. This we will now proceed to do.
%where $V$ is (as in our Viscous Froth simulations) the shearing velocity. 

\begin{figure}[htbp]
\begin{center}
\includegraphics[height=5cm,width=9cm]{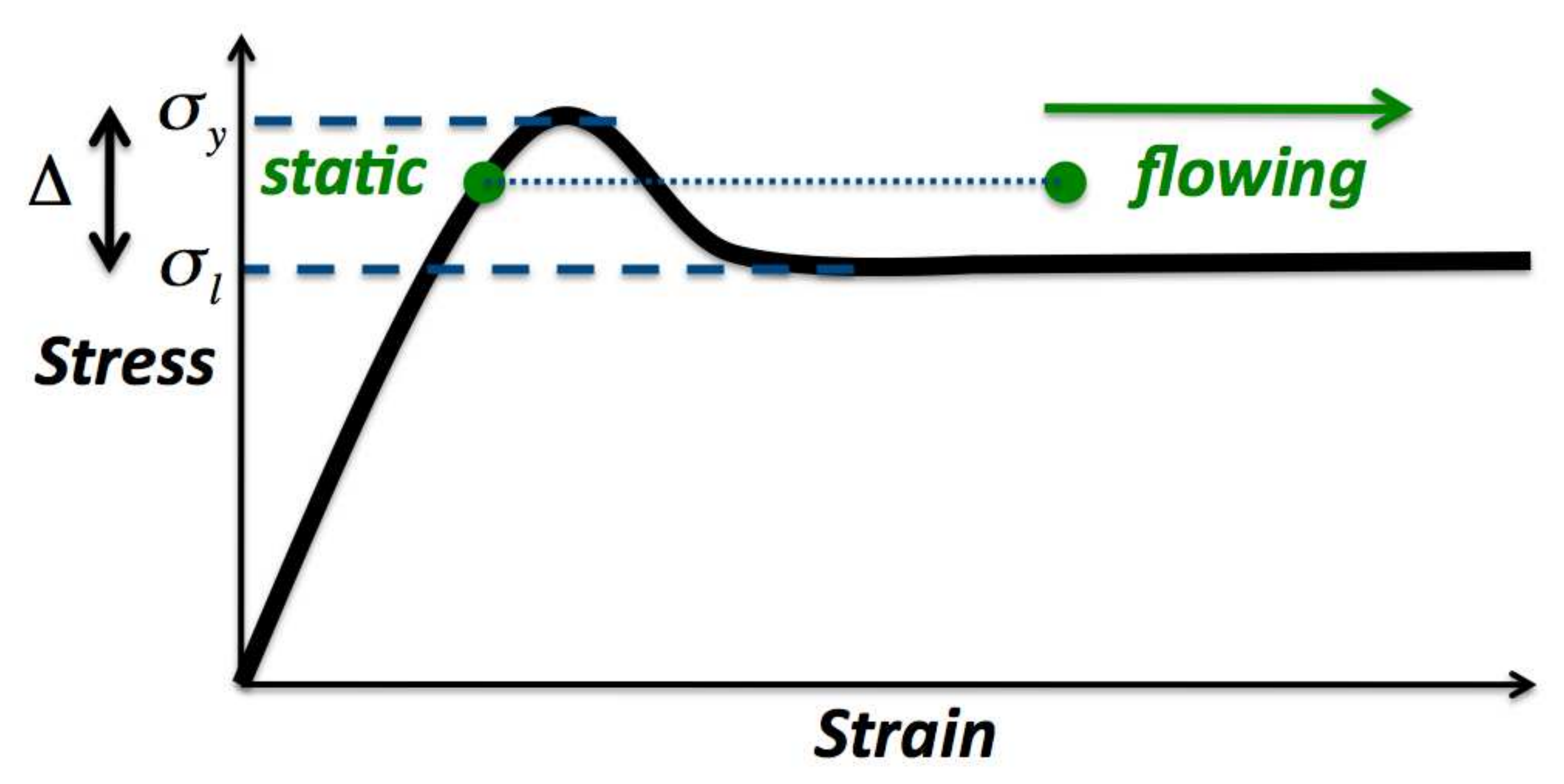} \hfil
\end{center}
\caption{An illustration of a stress vs strain relation incorporating the idea of distinct yield and limit stresses, denoted by $\sigma_{y}$ and $\sigma_{l}$ respectively. The filled circles indicate that the foam can co-exist at the same stress at the boundary between flowing and non-flowing regions.}
\label{f:continuum-hump}
\end{figure}

In a recent paper \cite{Weaire:09}, Weaire \emph{et al.} introduced the idea of distinct yield $\sigma_{y}$ and limit $\sigma_{l}$ stresses as a possible mechanism for localisation in the absence of viscous drag. An illustration of the typical stress vs strain picture is shown in Fig. \ref{f:continuum-hump}. The constitutive stress relation thus becomes

\begin{equation}
 \sigma=\sigma_{l}+c_{v}\dot{\epsilon}^a
 \label{eq:hb-new}
\end{equation}
\noindent
where $\sigma_l$ denotes the limit stress. When the magnitude of the stress overshoot, $\Delta=\sigma_y-\sigma_l$ is set to zero we recover the original Herschel-Bulkley relation (see eq. \ref{eq:hb-relation}).

If shear localisation is present in a foam, there exists at least one point $y_B$ which lies on the boundary between flowing and stationary regions. In our VF simulations, this corresponds to the point at which the velocity profile intercepts the $v=0$ axis (see, for example Fig. \ref{f:poly-profile-drag}). At this point, the system can co-exist at the same value of stress in both static and flowing regions, as indicated by the filled dots in Fig. \ref{f:continuum-hump}. The stress at $y_B$ can take on any value between $\sigma_{l}$ and $\sigma_{y}$ as the foam is sheared. From eq. \ref{eq:hb-new} we see that this leads to the inequality

\begin{equation}
0\le c_{v}\dot{\epsilon}(y_B)^{a}\le\Delta \ .
\label{eq:continuum-inequality}
\end{equation}
%where the stress overshoot, $\Delta=\sigma_{y}-\sigma_{l}$. 
\noindent 
As $V$ is increased, \emph{on average} we expect the viscous stress $c_v\dot{\epsilon}(y)^a$ between $0$ and $y_B$ to cause the stress in the flowing region to lie closer to $\sigma_{y}$ so that the stress overshoot is less evident. However, at low $V$, the effect is obvious (see Fig. \ref{f:qs-stress}) and may have important effects.

The differential equation given by eq. \ref{eq:diff-equation} may be solved numerically, yielding velocity profile solutions of the kind we envisage, which are valid if they satisfy the inequality given by eq. \ref{eq:continuum-inequality}. As the local strain rate is defined as
\begin{equation}
\dot{\epsilon}(y)=\left|\frac{dv(y)}{dy}\right|\ ,
\label{eq:local-strainrate}
\end{equation}
\noindent
in this case \cite{WeaireBarryHutzler:09}, the quantity $\dot{\epsilon}(y_B)$ can be directly measured from the calculated velocity profiles, provided they intersect the $v=0$ axis at some finite value. 

\begin{figure}[htbp]
\begin{center}
\includegraphics[width=9cm]{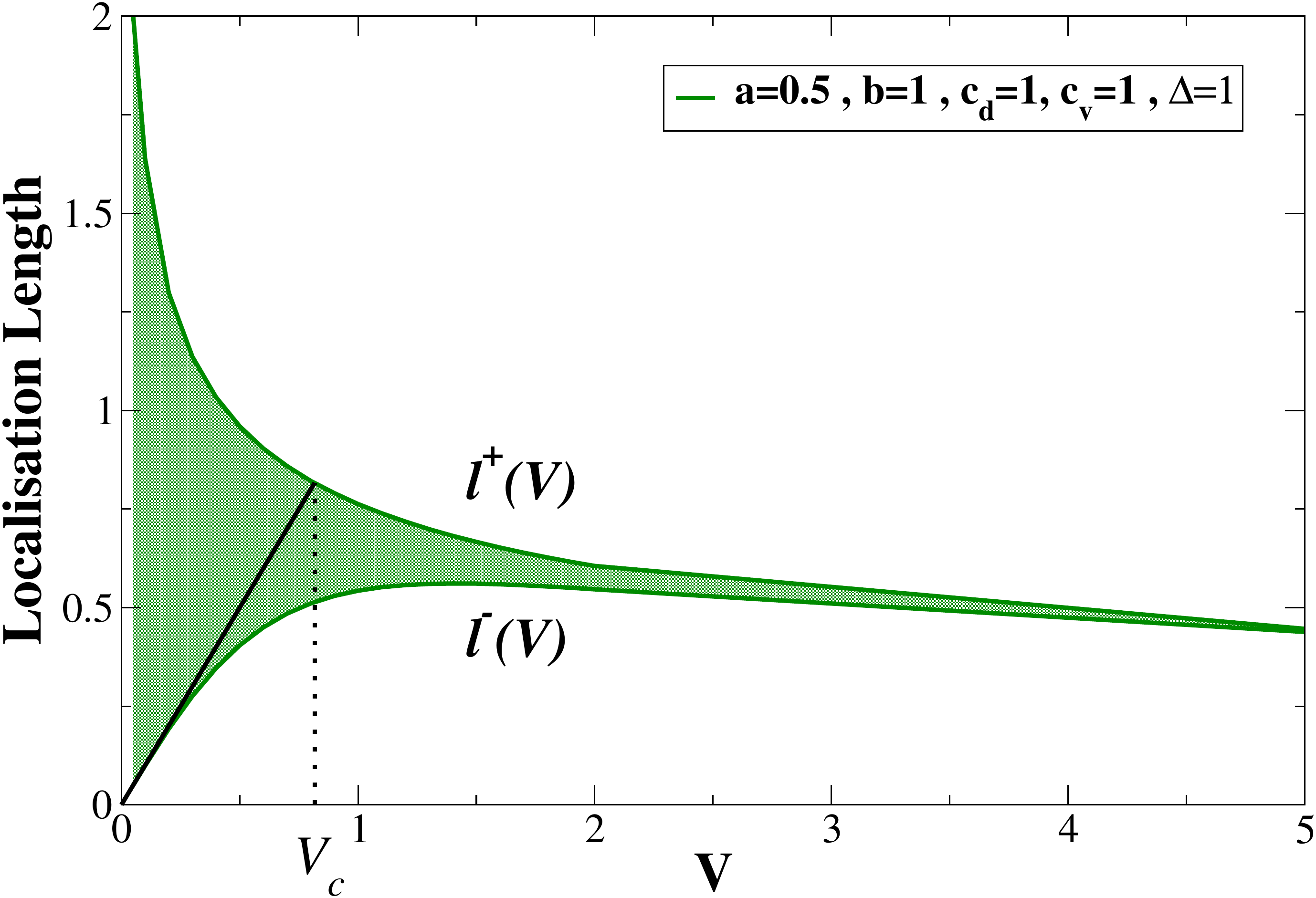} \hfil
\end{center}
\caption{Results from a numerical solution of the Continuum Model incorporating the inequality given by eq. \ref{eq:continuum-inequality}. For low $V$, there is a large range of possibilities for localisation length, while for high $V$, the range of allowed lengths converges to a narrow band. $V_c$ is our critical velocity below which a large range of localisation lengths is possible; see eq. \ref{eq:critical-velocity}. $l^{+}(V)$ and $l^{-}(V)$ denote the upper and lower bounds to the range of allowable solutions, respectively.}
\label{f:continuum-scatter}
\end{figure}

The results of these calculations can be seen in Fig. \ref{f:continuum-scatter}, where we have solved the model numerically for the values $a=0.5$, $b=c_{d}=c_{v}=\Delta=1$. The upper bound ${l^{+}}(V)$ corresponds to where the shear stress $\sigma(y_B)=\sigma_l$, where the viscous stress $c_{v}\dot{\epsilon}(y_B)^{a}=0$ and the analytic solution for localisation length given by eq. \ref{eq:continuum-localisation}. The lower bound ${l^{-}}(V)$ corresponds to where the shear stress $\sigma(y_B)=\sigma_y$ and where the viscous stress $c_{v}\dot{\epsilon}(y_B)^{a}=\Delta$. Thus, for a given $V$, $l^{-}(V)\le l(V) \le l^{+}(V)$ gives the range of allowed solutions, indicated by the shaded region in Fig. \ref{f:continuum-scatter}. 
%A more detailed description of this procedure can be found in \cite{WeaireBarryHutzler:09}.

We note that Fig. \ref{f:continuum-scatter} is qualitatively similar to Fig. \ref{f:poly-localisation-velocity}, with a large range of possible localisation lengths at low $V$ and convergent behaviour at high $V$. Remarkably, the model predicts that for low $V$, the localisation length can take any value $0<l<\infty$. This prediction, of course, holds only in the presence of viscous drag.

An important question to be answered is how does one define the critical velocity $V_{c}$ below which the foam can take on a wide range of localisation lengths? If one assumes that as $V\to0$, the velocity profile becomes approximately linear, then $\dot{\epsilon}=\frac{V}{l}$, where $l$ is the localisation length. If we are on the lower bound, from eq. \ref{eq:continuum-inequality} we see that $\Delta=c_{v}\left(\frac{V}{l}\right)^{a}$, or expressing it in a more convenient form,

\begin{equation}
l=\left(\frac{c_{v}}{\Delta}\right)^{\frac{1}{a}}V\ .
\label{eq:l-lower}
\end{equation}

We are interested in the point where this line intersects the upper bound $l^{+}(V)$, which is given by eq. \ref{eq:continuum-localisation}. We solve this pair of simultaneous equations (eq. \ref{eq:continuum-localisation}, \ref{eq:l-lower}) in terms of $V$ and choose to define the point of intersection as our critical velocity $V_{c}$ (see Fig. \ref{f:continuum-scatter}). This yields

\begin{equation}
V_{c}=\Delta^\frac{a+1}{a(1+b)}\left(\frac{a(1+b)}{1+a}\right)^{\frac{1}{1+b}}\left(\frac{1}{c_{d}(c_{v})^{\frac{1}{a}}}\right)^{\frac{1}{1+b}}\ .
\label{eq:critical-velocity}
\end{equation}
\noindent
 
To make a more quantitative comparison between continuum theory and the VF results presented in Sec. \ref{sec:details}, a more detailed study of the relationship between the parameters of both models is required. We present such a study in the next section.

In one of the earliest publications on this subject, Kabla \& Debregeas \cite{Kabla:03} attribute shear localisation in quasi-statics to what they call `self amplification'. This idea is qualitatively the same as the ideas presented in this section. This approach may have the capacity to explain other results in the literature, particularly \cite{Wang:07} where shearing experiments are performed for an ordinary Bragg raft (where there are no confining plates) in a straight geometry. In these experiments (as in our VF simulations) variations in the averaged velocity profiles are observed between experiments but averages over several experiments converge much better.

\section{Relating the Continuum Model to the Viscous Froth Model}
\label{sec:multiscale}

We now proceed to relate the parameters of the (microscopic) VF Model and the (macroscopic) Continuum Model. This is done using a combination of numerical and analytic approximations.

%In Sec. \ref{sec:stress} we perform \emph{quasi-static} simulations to demonstrate and measure the magnitude of the stress overshoot. In Sec. analytic approximations and dimensional argument

%In the absence of viscous drag, this theory does not exclude the possibility of shear-banding occuring away from the moving boundary. In one VF simulation, for $\lambda=0.05$ and $V=1$ we have observed such behaviour. After a transient, a shear band forms in the centre of the foam sample in a manner reminiscent of quasi-static results by Wyn \emph{et al.} \cite{Wyn:08}. We have not of yet, however, been able to reproduce this behaviour with other foam samples.

%\subsection{The stress overshoot}
%\label{sec:stress}

\begin{figure}[htbp]
\begin{center}
\includegraphics[width=9cm]{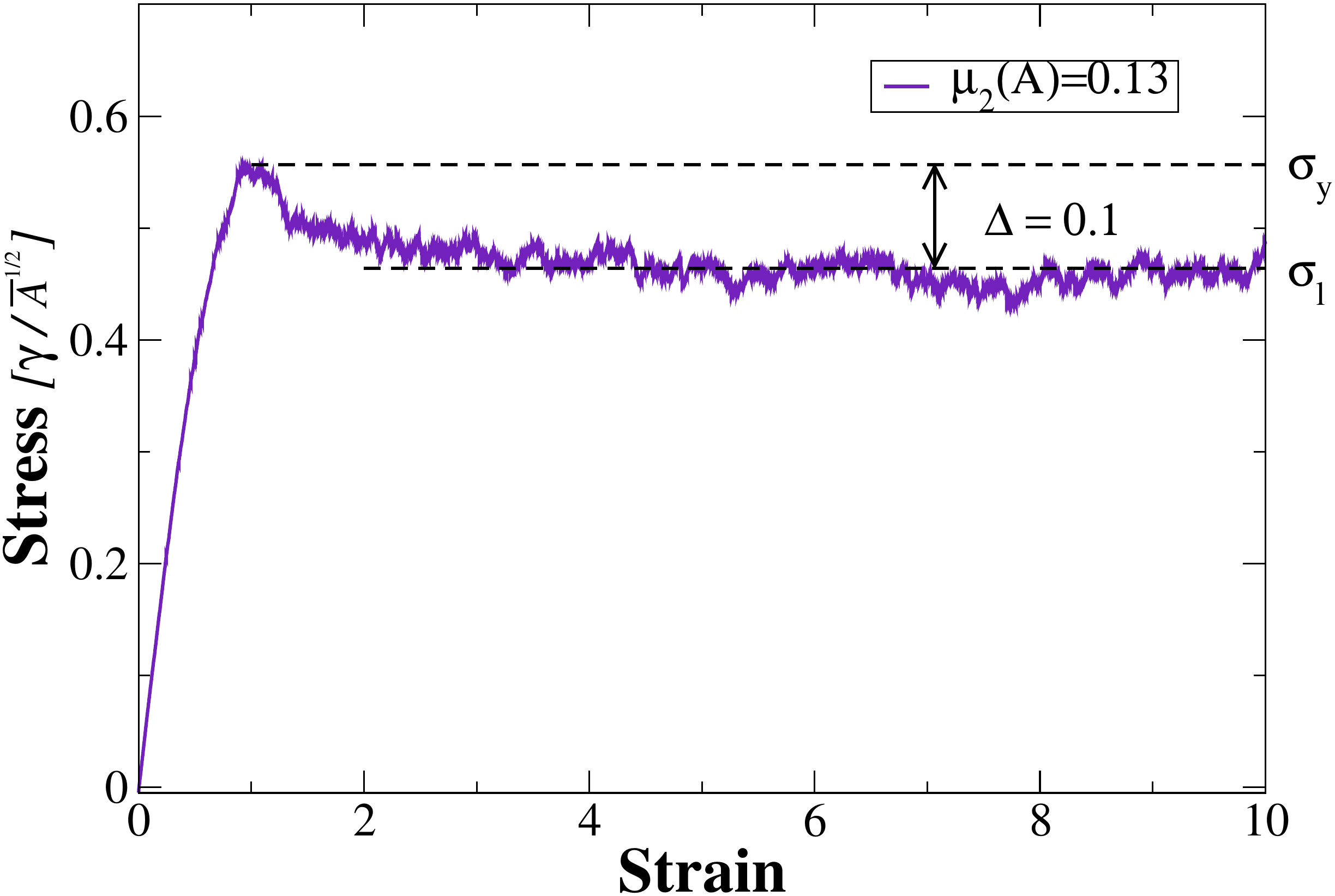} \hfil
\end{center}
\caption{Averaged shear stress data for 29 foam samples of disorder $\mu_{2}(A)=0.13\pm0.03$. The stress overshoot is clearly evident. The yield stress $\sigma_{y}$ is taken to be the maximum stress value, which occurs at a strain of unity. The limit stress $\sigma_{l}$ is the stress average from a strain of 2 to 10. Calculation performed using \emph{quasi-static} simulations.}
\label{f:qs-stress}
\end{figure}

To demonstrate preliminary evidence of existence of the stress overshoot in simulation, we have performed quasi-static calculations using the Surface Evolver (of the type mentioned in in Sec. \ref{intro}) for 29 foam samples of disorder $\mu_{2}(A)=0.13\pm0.03$ with $N_b=100$ bubbles. This effectively sets the viscous stress $c_v\dot{\epsilon}(y)^a$ to zero, thereby allowing us to obtain an accurate estimate of the magnitude of the stress overshoot, $\Delta$. The foam samples are created using the process outlined in Sec. \ref{sec:creation}. The shear stress $\sigma_{xy}$ (defined in \cite{Kraynik:03}) is recorded for each simulation and subsequently averaged; see Fig. \ref{f:qs-stress}. 

\begin{figure}[htbp]
\begin{center}
\includegraphics[width=9cm]{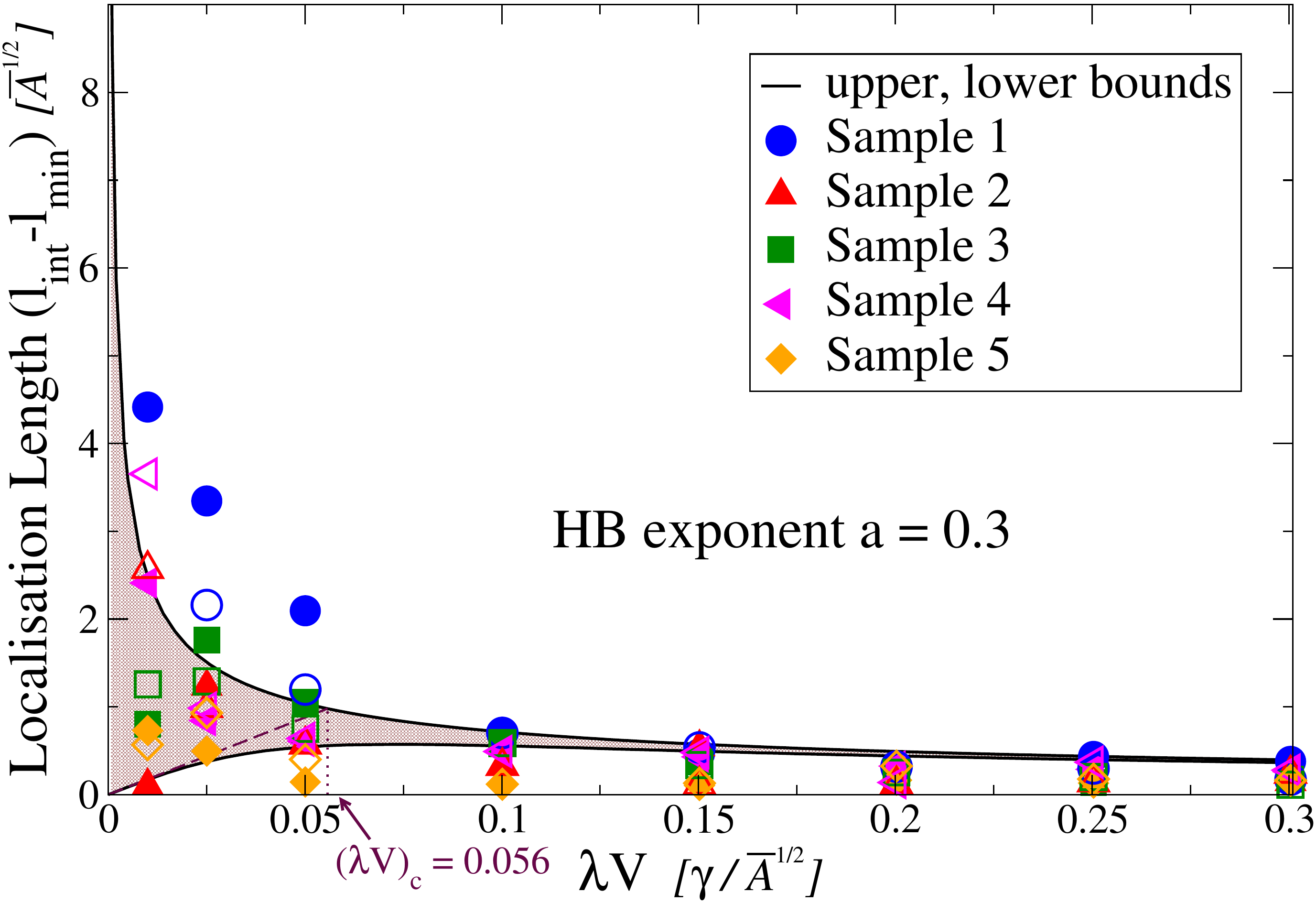} \hfil
\end{center}
\caption{A comparison of the VF simulation results (open and filled symbols) and the prediction for the range of allowed localisation lengths as given by the Continuum Model (shaded region). A Herschel-Bulkley exponent of $a=0.3$ and a stress overshoot of $20\ \%$ is found to give a good fit to the data. The critical cross-over point, $(\lambda V)_c$, as given by eq. \ref{eq:lv-crit} indicates the point below which the system yields a wide range of localisation lengths. }
\label{f:prediction}
\end{figure}

As there is localisation in these simulations (at either the moving or stationary boundary) which affects our stress measurements, the limit stress $\sigma_{l}$ reported here must be treated as an approximate measurement. The value of the yield stress $\sigma_{y}$, however, is exact, as up to a strain of unity, the foam is in the elastic regime and the bubble motion has not yet localised. We measure the magnitude of the stress overshoot to be $\Delta=0.1\ \gamma / \bar{A}^{1/2}$, which corresponds to a $17\%$ overshoot. In the calculation shown in Fig. \ref{f:prediction}, a $20\%$ overshoot is used.

%\subsection{The wall drag force}

The drag force per unit area for the Continuum Model is given by eq. \ref{eq:drag-force} and acts in the direction of shear. We wish to relate this to the the drag force of the VF Model, $\lambda v^\perp$, which is a force per length and acts in the normal direction to a soap film (see Fig. \ref{f:VF_forces}). Trivially, the drag exponent, $b=1$. The numerical prefactor $c_d$ may be calculated analytically for a 2D hexagonal honeycomb structure, which serves as a reasonable approximation. We also take into account the direction in which the drag force is defined and the orientation of soap films in the foam.

The drag force per unit area must be proportional to the total length of the soap films in that area. For the honeycomb, this yields

\begin{equation}
 c_d\propto\sqrt{\frac{2\sqrt{3}}{\bar{A}}}\ .
\end{equation}

\begin{figure}[htbp]
\begin{center}
 \includegraphics[width=9cm]{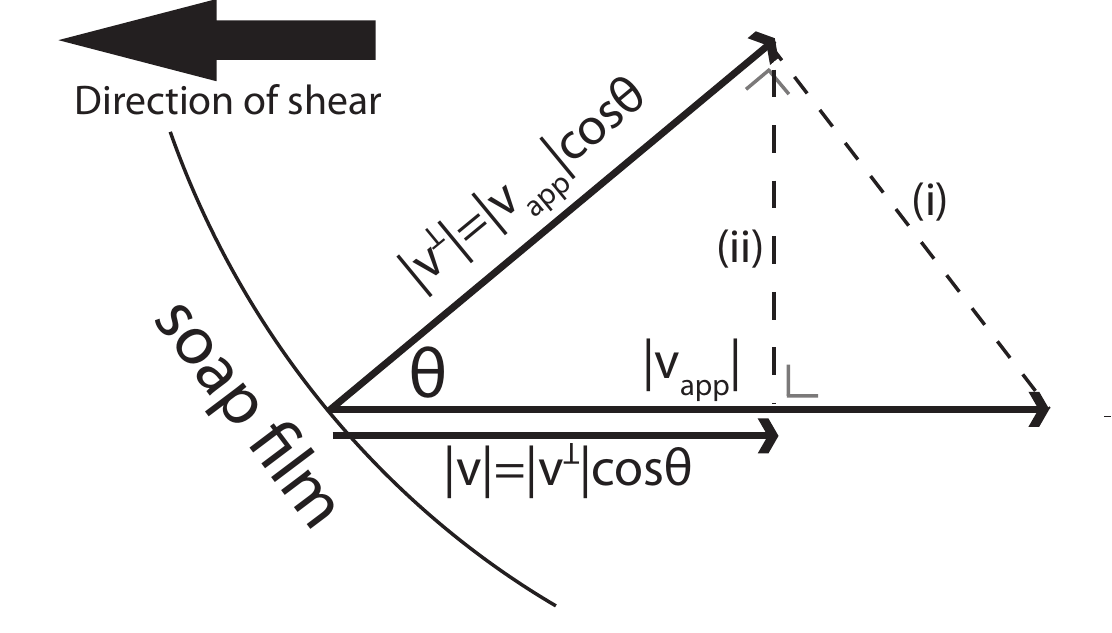}\hfil
\end{center}
\caption{
Two projections are necessary to relate the velocity $\vec{v_{app}}$ of a soap film segment in the VF model to the local velocity \vec{v} of the Continuum Model:
(i) projection of the average velocity of a soap film segment $\vec{v_{app}}$ in the direction of the normal to that segment, and (ii) projection of the normal velocity of the soap film segment $\vec{v^\perp}$ back in the direction of shear. $\theta$ is the angle between the normal vector to the soap film segment and the shear direction. It is the magnitude of the vectors that is displayed in the figure.}
\label{f:projection}
\end{figure}

%\noindent
In the VF simulations, it is observed that bubbles move on average only in the direction of shear. The magnitude of this `apparent' velocity is denoted by $v_{app}$ in Fig. \ref{f:projection}. However, the drag force for the VF model \emph{by definition} points in the direction of the normal to a soap film, and so we project $\vec{v_{app}}$ in this direction (see Fig. \ref{f:projection}(i)). This results in $|\vec{v^\perp}|=|\vec{v_{app}}| \cos\ \theta$, where $\theta$ is the relative angle between the normal vector to the soap film and the shear direction. To relate the normal drag force to the actual drag force of the Continuum Model, we need to project $\vec{v^\perp}$ in the shear direction (see Fig. \ref{f:projection}(ii)), resulting in $|\vec{v}|=|\vec{v^\perp}| \cos \theta=|\vec{v_{app}}|(\cos\ \theta)^2$.

Finally, we consider how the orientation of the soap films in our foam might affect the drag force. We assume that the foam is isotropic and proceed to average over all possible values of $\theta$. As $<(\cos\ \theta)^2>=1/2$, our final expression for the continuum drag force coefficient $c_d$ is

\begin{equation}
 c_d  = \hat{c_d} \lambda =  \frac{1}{2}\sqrt{\frac{2\sqrt{3}}{\bar{A}}}\lambda \ ,
 \label{eq:cd}
\end{equation} 
\noindent
giving the (continuum) drag force the required dimensions of force per area.
%In Viscous Froth, the drag exponent, $b=1$ \cite{Kern:04}. The drag constants from the Continuum Model, $c_{d}$ (dimensions $\frac{FT}{L^3}$), and the VF Model, $\lambda$ (dimensions $\frac{FT}{L^2}$), are related by
%\begin{equation}
% c_d=\frac{1}{2}\sqrt{\frac{2\sqrt{3}}{\bar{A}}}\lambda
%\end{equation}
%\noindent
%where the numerical pre-factor is calculated for a hexagonal foam structure. 

%\subsection{The viscous stress}
In Sec. \ref{sec:details}, we showed how the viscous stress has a $\lambda V$ dependence using dimensional arguments (see eq. \ref{eq:viscous-stress}). Using these arguments, but rather defining the strain rate as a locally changing quantity (see eq. \ref{eq:local-strainrate}), we see that

\begin{equation}
c_v=\hat{c_v}\gamma^{1-a}{\bar{A}}^{a-1/2}\lambda^a
\label{eq:cv}
\end{equation}
\noindent
where the Herschel-Bulkley exponent $a$ and the dimensionless quantity $\hat{c_v}$ are free parameters.

Using all of the approximations calculated in this section, we proceed to solve eq. \ref{eq:diff-equation} numerically, accepting solutions only if they obey the inequality given by eq. \ref{eq:continuum-inequality}, as done in Sec. \ref{sec:Continuum}. The key difference here is that localisation length is a function of the product $\lambda V$. 

The upper bound for the Continuum Model prediction is formulated in terms of $\lambda V$ by inserting eq. \ref{eq:cd} and eq. \ref{eq:cv} into eq. \ref{eq:continuum-localisation}, resulting in

\begin{equation}
l^+(\lambda V) = \left( \frac{2 a \hat{c_v} \sigma^{1-a} {\bar{A}}^{a-1/2} }{(1+a) \hat{c_d}}   \right)^\frac{1}{1+a} (\lambda V)^\frac{a-1}{1+a}\ .
\label{eq:lv-upper}
\end{equation}
\noindent
The corresponding lower bound must be found numerically. To simplify this calculation, we fix $\lambda$ and allow $V$ to vary.

A comparison of the VF and Continuum Model results can be seen in Fig. \ref{f:prediction}, where values of $a=0.3$ and $\hat{c_v}=0.26$ are chosen as they give a reasonable prediction for both upper and lower bounds (although $0.2<a<0.4$ gives a reasonable fit to the upper bound). The shaded region between these bounds indicates the range of all allowable localisation lengths, as predicted by the Continuum Model. Filled and open symbols represent the VF simulation results, which are the same as in Fig. \ref{f:poly-localisation-velocity}, only with the minimum localisation length of $l_{min}={\bar{A}}^{1/2}$ subtracted to coincide with the Continuum Model predictions which give $l=0$ for $V\to\infty$.

%subtracted in order to satisfy the second boundary condition associated with eq. \ref{eq:diff-equation}. 
%That the HB exponent of $a=0.3$ is similar to the experimentally reported value of $a=0.36$ by Katgert \emph{et al.} \cite{Katgert:08,Katgert:09} is curious given that the forces on individual bubbles/films are quite different in both cases. The local drag forces in the VF model are linear, whereas in these experiments, \emph{nonlinear} local drag laws for bubble-bubble and bubble-wall interactions are reported. 

The definition for the critical cross-over point, as given by eq. \ref{eq:critical-velocity} may also be formulated in terms of $\lambda V$. This is achieved by inserting eq. \ref{eq:cv} into eq. \ref{eq:l-lower} and finding the point at which this line intersects eq. \ref{eq:lv-upper}. Alternatively, one may insert eq. \ref{eq:cd} and eq. \ref{eq:cv} into eq. \ref{eq:critical-velocity}. This gives
\begin{equation}
(\lambda V)_c = \sqrt{  \frac{2a\Delta^{\frac{1+a}{a}} (  \hat{c_v} \sigma^{1-a} {\bar{A}}^{a-1/2}   )^{-{1}/{a}}   }{(1+a) \hat{c_d}}  }\ ,
\label{eq:lv-crit}
\end{equation}
\noindent
which is illustrated by the dotted lines in Fig. \ref{f:prediction}. The calculated critical cross-over point $(\lambda V)_c=0.056\ \gamma/ \bar{A}^{1/2}$ fulfills its promise of offering a reasonable estimate of the point below which the system yields a wide range of localisation lengths.

While the comparison between the  VF Model and the Continuum Model presented in this section gives a fascinating theoretical explanation for the simulation results discussed, its details are far from precise. The location of the upper bound in Fig. \ref{f:prediction} is simply an estimate, and further simulations may be needed to determine its exact location. In addition, many approximations were taken in relating the parameters of the two models. However, it is remarkable that \emph{despite} these approximations, a robust prediction can still me made.

\section{Outlook}

The apparent agreement of the simulation results in this paper with published experimental work suggests that the 2D VF Model may have further potential for describing realistic foam dynamics. For more detailed studies to be conducted, however, the VF algorithm will need to be improved to decrease the required computation time for these types of simulations. Issues that we will address include the effect of $\mu_2(A)$ on localisation with the 2D VF Model and on the value of the HB exponent. In addition, the dependence of the magnitude of the stress overshoot $\Delta$ on $\mu_{2}(A)$ will be investigated as it is critical to our understanding of its role as a mechanism for shear localisation. It will be of interest to observe what happens to the location of the shear-band for samples with higher $\mu_2(A)$, in light of the results published in \cite{Wyn:08}. We also intend to compare our VF simulations with simulations using the Soft Disk Model \cite{Langlois:08}.

\section{Acknowledgements}
The author would like to acknowledge IRCSET Embark for funding this project. IITAC, the HEA, the National Development Plan and the Trinity Centre for High Performance Computing are acknowledged for the use of the computing facilities at TCD. S.J. Cox, Aberystwyth is thanked for his useful input and correspondence in relation to this work. This publication has emerged from research conducted with the financial support of the European Space Agency (MAP AO-99-108:C14914/02/NL/SH and AO-99-075:C14308/00/NL/SH). This material is based upon works supported by the Science Foundation Ireland under Grant No. (08/RFP/MTR1083  and STTF 08). We would also like to thank the anonymous referees whose input helped greatly in improving the quality of this manuscript.

% BibTeX users please use
%\bibliographystyle{svjour}
%\bibliography{VF_paper}
%\begin{thebibliography}{}
%Author, Journal \textbf{Volume,} (year) page numbers.
%Author, \textit{Book title} (Publisher, place year) page numbers
% etc
%\end{thebibliography}

\end{document}